\documentstyle[12pt]{article}
\newcommand{\beq}{\begin{equation}}
\newcommand{\eeq}{\end{equation}}
\newcommand{\beqa}{\begin{eqnarray}}
\newcommand{\eeqa}{\end{eqnarray}}
\newcommand{\beqar}{\begin{eqnarray*}}
\newcommand{\eeqar}{\end{eqnarray*}}
\newcommand{\al}{\alpha}
\newcommand{\dd}{{\rm d}}
\newcommand{\eps}{\epsilon}
\newcommand{\Ga}{\Gamma}
\newcommand{\ka}{\kappa}
\newcommand{\calh}{{\cal H}}

\newcommand{\eg}{{\it e.g.,}\ }
\newcommand{\ie}{{\it i.e.,}\ }
\newcommand{\Tr}{{\rm Tr}}
\newcommand\prt{\partial}
\newcommand\h{H}
\newcommand\hh{h}
\newcommand{\htil}{\widetilde{H}}
\newcommand{\hhtil}{\tilde{h}}
\newcommand{\abar}{{\bar{a}}}
\newcommand{\bbar}{{\bar{b}}}
\newcommand{\cbar}{{\bar{c}}}
\newcommand{\zbar}{{\bar{z}}}
\newcommand{\nbar}{{\bar{n}}}
\newcommand{\qtil}{\tilde{q}}
\newcommand{\calm}{{\cal M}}
\newcommand{\vareps}{\varepsilon}
\newcommand{\one}{1\!\!1}

\begin{document}
\begin{titlepage}
\rightline{\small hep-th/9705079 \hfill McGill/97-08}
\vskip 5em

\begin{center}
{\bf \huge
Hermitian D-brane solutions}
\vskip 3em

{\large G. Michaud\footnote{email:
gmichaud@hep.physics.mcgill.ca\hfil}} and
{\large R.C. Myers\footnote{email: rcm@hep.physics.mcgill.ca\hfil}}
\vskip 1em

{\em	Department of Physics, McGill University \\
        Ernest Rutherford Physics Building\\
	Montr\'eal, Qu\'ebec, Canada H3A 2T8}
\vskip 4em

\begin{abstract}
A low-energy background field solution describing D-membrane configurations
is constructed which is distinguished by the appearance of a Hermitian
metric on the internal space. This metric is composed of
a number of independent harmonic functions on the
transverse space. Thus this construction generalizes the usual 
harmonic superposition rule. The BPS bound of 
these solutions is shown to be saturated indicating that they are
supersymmetric.
By means of T-duality, we construct more solutions
of the IIA and IIB theories.
\end{abstract}
\end{center}

\end{titlepage}

\setcounter{footnote}{0}
\section{Introduction}

In the past two years, evidence has accumulated which indicates
that all five superstring theories are in fact perturbative
expansions about different points in the phase space of some
more fundamental
theory, the so-called M-theory \cite{drama}. In particular, 
this interpretation is suggested by the success in relating
these theories by various duality transformations. This
progress has been marked by the realization that
extended objects, other than just strings, 
play a very important role. Of note in the
type IIA and IIB superstring theories are 
Dirichlet branes (D-branes) which have been recognized as the 
carriers of Ramond-Ramond (RR) charges \cite{Polchin} -- see also
\cite{Polchin2}.

One of the remarkable aspects of D-branes is that they can be
analysed as external sources in the framework of perturbative
worldsheet calculations. When the effective coupling
between the strings and D-branes  grows, a complementary description
of the D-branes is provided by solutions of the effective low energy
supergravity equations of motion. Investigating both of these approaches
to D-brane physics has proven useful. In particular, both methods
converged in recent calculations of black hole entropy \cite{blackholes}.
In this analysis, perturbative techniques were used in making a counting
of the ground state degeneracy of certain D-brane bound states, while
the corresponding low energy solutions provided a black hole background
for which one could calculate the usual Hawking-Bekenstein entropy.
The precise matching of the results in these two regimes
has provided some striking new insights into
the underlying microscopic degrees of freedom for (at least near-extremal)
black holes.

One aspect of D-brane physics which has recently attracted some 
attention \cite{angledrama2,angledrama1,angles,nour,mir}
is the possibility of constructing marginally
bound states in which the D-branes intersect at
angles \cite{Berkooz}, other than $0$ and $\pi/2$. In these
configurations, 
the various D-branes are typically related by an $SU(N)$ subgroup of 
rotations, in order to preserve supersymmetry --- actually 
\cite{angledrama2,gaunt}
indicates that other subgroups are possible.
An explicit background field solution describing several D-membranes rotated
by $SU(2)$ rotations was presented in \cite{angles}.
The general form of this solution was complicated and did not appear
simply related to any previously known solutions.

In this paper, we will show that this solution 
with angled D-membranes falls within a class
of D-brane solutions characterized by a Hermitian metric on
the internal space. This metric contains a number of independent
harmonic functions, and so the new solution represents a
generalization of the usual harmonic superposition construction \cite{ortho}.
These solutions, presented in section 2, describe a general class 
of six-dimensional D-membrane configurations.
In section 3, we provide a similar class of four-dimensional solutions
describing configurations of D-membranes, as well as
D6-branes, wrapped around the internal space\footnote{While this
work was being completed, a paper by Balasubramanian, Larsen and 
Leigh \cite{Bala} appeared which describes this solution in terms
of D-membranes at angles.}. For these solutions,
the metric on the internal six-torus is again Hermitian.
We also explicitly show that this solution
saturates the BPS bound, and preserves one-eighth of the supersymmetries.
The following section is devoted
to building some simple extensions of this Hermitian metric solution
through T-duality transformations.  We conclude with a brief discussion of 
the results in section \ref{discuss}.
Our notation and conventions follow those established in \cite{us}.

\section{Six-dimensional solutions}

Ref.~\cite{angles} presented a solution describing an arbitrary number of
D-membranes oriented at angles with respect to one another. There, this
solution was derived directly by solving the low-energy
supergravity equations of motion --- see also \cite{nour}.
Later, the two membrane solution was also reproduced through
a series of boosts and duality transformations applied to
two orthogonal M-branes \cite{mir}. As presented in \cite{angles},
the solution took a complicated form in terms of
harmonic functions $X_a$ and rotation angles $\al_a$, where
the index $a$ is labeling the $a$'th
D-membrane.  In particular
following the analysis of ref.~\cite{Berkooz},
the membrane rotations were described as a particular set of
SU(2) transformations, however no use of complex coordinates was
made in presenting the solution. One finds however that the latter
can be employed to greatly simplify, as well as generalize, this
solution.

We begin by defining complex coordinates on the
effective worldvolume\footnote{We 
include any directions in which the D-branes are delocalized as
world-volume coordinates. We will refer to all of the spatial
(effective) world-volume coordinates with $y^i$, while the transverse
coordinates will be $x^i$.}:
\beq
z^1 = {1\over\sqrt{2}}\left(y^1 +iy^2\right) 
\qquad\mbox{and}\qquad 
z^2 = {1\over\sqrt{2}}\left(y^3 +iy^4\right)
\eeq
together with their complex conjugates, $\zbar^1$ and $\zbar^2$.  We then 
have the solution
\beqa
ds^2 &=& \sqrt{\calh}\left( {-dt^2 +2 \,\h_{a\bbar}\, dz^a d\zbar^b\over \calh} 
+\sum_{i=5}^9 (dx^i)^2\right) \nonumber\\
A^{(3)} &=& \pm{i\over \calh} \h_{a\bbar}\; dt\wedge dz^a\wedge d\zbar^b
\nonumber\\
e^{2\phi} &=& \calh^{{1\over 2}}\label{oldsolution}
\eeqa
where $\h_{a\bbar}$ is a Hermitian matrix given by 
\beq
\h_{a\bbar} =\left(
\begin{array}{cc}
1+A & C \\
C^* & 1+B
\end{array}
\right) 
\label{matrix}
\eeq
and $\calh^2 =\det \h=(1+A)(1+B)-|C|^2$.
The entries of $\h_{a\bbar}$ are independent harmonic functions 
of the $x^i$'s. That is, they solve the flat space Poisson's
equation in the transverse space, \eg $\delta^{ij}\partial_i\partial_j
A = {sources}$. Implicitly here, we assume that these
harmonic functions all vanish asymptotically leaving $\h_{a\bbar}\simeq
\delta_{a\bbar}$. At the same time, we think 
of the $y^i$ coordinates as being compact with range $2\pi L_i$, 
and hence asymptotically the internal space is simply a square four-torus.
One can easily change the asymptotic moduli of
this internal space by either modifying these periodicities
or by adding constants to the harmonic functions in eq.~(\ref{matrix}).
One must demand,
however, that $\h_{a\bbar}$ remains invertible. We also note that
the combination
\beqa
\det(\h)\,\h^{\abar b}=
\left(
\begin{array}{cc}
1+B & -C \\
-C^* & 1+A
\end{array}
\right)
\label{invert}
\eeqa
takes a simple form in terms of the harmonic functions. Here,
$\h^{\abar b}$ denotes the inverse of the Hermitian metric.

If we choose zero for the (complex)
off-diagonal component of $\h_{a\bbar}$, \ie $C=0$,
then the above solution simply reduces
to that describing orthogonal D-membranes, constructed with the usual
harmonic superposition rule \cite{ortho}.
To relate the above solution to that of angled D-membranes \cite{angles},
one makes the following choice for the harmonic
functions\footnote{Actually this only reproduces the solution of \cite{angles}
up to an interchange of the internal coordinates, $y^2\leftrightarrow y^3$,
and a gauge transformation of the RR potential.}:
\beqa
A &=& \sum_{a=1}^n X_a \cos^2\alpha_a \nonumber\\
B &=& \sum_{a=1}^n X_a \sin^2\alpha_a \nonumber\\
C=C^* &=& \sum_{a=1}^n X_a \sin\alpha_a\cos\alpha_a 
\nonumber
\eeqa
In this case, the rotations acting on the individual membranes can
be described as SU(2) transformations: $z^b\rightarrow 
\left[U_a\right]^b{}_c\,z^c=\left[\exp(i\alpha_a\sigma_2)\right]^b{}_c\, z^c$.
Our Hermitian metric can then be
written as
\beq
\h_{b\cbar}=\delta_{b\cbar}+\sum_{a=1}^n X_a\,\left[U_a\right]^1{}_b
\,\left[U_a^*\right]^1{}_\cbar\ \ .
\label{general6}
\eeq
Written in this form, we can easily generalize the solution by allowing the
matrices $U_a$ to be arbitrary SU(2) transformations.

Following \cite{angles},
it is straightforward to calculate the ADM mass and charge densities
for the general solution, and further to show that these saturate
the BPS bound. Hence these solutions preserve one-quarter of the
supersymmetries. We leave the details
of these calculations for the next section, where the
four-dimensional solution is discussed.

It is interesting to consider the special case where all of the
harmonic functions have a common centers\footnote{We are grateful
to A.A. Tseytlin for interesting discussions on this point.}.
With a single center, the Hermitian metric takes the form
\beq
\h_{a\bbar}=\delta_{a\bbar}+{\hh_{a\bbar}\over|\vec{x}-\vec{x}_o|^3}
\label{single}
\eeq
where $\hh_{a\bbar}$ is a {\it constant} Hermitian matrix. In this
case, however, it will always be possible to find an SU(2) transformation
which diagonalizes the latter, \eg
\beq
\hhtil=U\,\hh\,U^\dagger=
\left(
\begin{array}{cc}
\hhtil_{11} & 0 \\
0 & \hhtil_{22}
\end{array}
\right)\ \ .
\label{diag}\eeq
Therefore by performing the (global) coordinate transformation $\tilde{z}^i
=U^i{}_j\,z^j$, the off-diagonal components of $\h_{a\bbar}$ will be
eliminated in which case the single-center solution is reduced
to that describing orthogonal D-membranes \cite{ortho}. Thus in the
special case where all of the centers for the harmonic functions
come together, our new Hermitian D-membrane solutions in fact only
reproduce the known
orthogonal membrane solution. These considerations may be
generalized to the special case of several common centers
\beq
\h_{a\bbar}=\delta_{a\bbar}+\sum_{i=1}^n
{\hh^{(i)}_{a\bbar}\over|\vec{x}-\vec{x}_i|^3}
\label{common}
\eeq
where $\hh^{(i)}_{a\bbar}= \mu_i\,\hh^{(1)}_{a\bbar}+
\rho_i\,\delta_{a\bbar}$ with constants $\mu_i$ and $\rho_i$ for
$i\ge 2$. Again, a coordinate
transformation can be found which will reduce this special
case of (\ref{oldsolution}) to the standard orthogonal
solution. Thus while generically our Hermitian solutions represent
new solutions, at many points in the moduli space they reduce to
known solutions \cite{ortho}.
Further, however, irrespective of the choice of the harmonic functions
in $\h_{a\bbar}$, as long as the centers all have finite separation
then the asymptotic fields will take the form given in eq.~(\ref{single}).
Therefore at long distances, physicists may always
interpret these general D-membrane configurations in terms of
orthogonal membranes.

\section{Four-dimensional solutions}

In this section, we will generalize the above solutions to the
case where the effective world-volume or the internal space occupies
six spatial dimensions leaving four noncompact directions.
Thus the D-membranes fill a six-torus described by
$y^i$ where $i=1,2,3,4,5,6$, while there are three transverse
spatial coordinates, $x^i$ with $i=7,8,9$.  
Since parallel D2- and D6-branes are supersymmetric \cite{Polchin2},
it is natural to extend these solutions to include D6-branes
wrapped around the six-torus.
Again, complex coordinates are useful and we define
\beq
z^a={1\over \sqrt{2}} \left(y^{2a-1} +i\,y^{2a}\right)
\eeq
for $a=1,2,3$. The natural generalization of eq.~(\ref{oldsolution})
is then\footnote{We have introduced an electric
seven-form potential to describe D6-branes. This represents a great
simplification over the usual magnetic one-form RR potential in
describing multi-center solutions. We also note here that
the solutions here and in the previous section are given in terms of
the string-frame metric.}
\beqa
ds^2 &=& \sqrt{\calh_2\calh_6} \left( {-dt^2 +2\,\h_{a\bbar}\,dz^a d\zbar^b
\over \calh_2\calh_6} +\sum^9_{i=7} (dx^i)^2 \right) \nonumber\\
A^{(3)} &=& \pm{i\over \calh_2} \h_{a\bbar}\; dt\wedge dz^a\wedge d\zbar^b 
\nonumber\\
A^{(7)} &=& \pm{i\over \calh_6}\; dt\wedge dz^1\wedge dz^2\wedge dz^3\wedge 
d\zbar^1\wedge d\zbar^2\wedge d\zbar^3 \nonumber\\
e^{2\phi} &=& {\calh_2^{1/2}\over \calh_6^{3/2}}\label{solution}
\eeqa
The Hermitian matrix $\h_{a\bbar}$ takes a slightly more complicated form
\beq
\h_{a\bbar}= \left(
\begin{array}{ccc}
(1+A)(1+D) -|E|^2 & C(1+D) + E^*F^* & F^* (1+A) +CE \\
C^* (1+D) +EF & (1+B)(1+D) -|F|^2 & E(1+B) +C^* F^* \\
F (1+A) +C^* E^* & E^* (1+B) +CF & (1+A)(1+B) -|C|^2
\end{array} \right) \; .
\label{hermit}
\eeq
and in this case
\beqa
\calh_2&=&\sqrt{\det \h}
\nonumber\\
&=&(1+A)(1+B)(1+D)-|C|^2(1+D)-|F|^2(1+A)-|E|^2(1+B) \nonumber \\
&&\qquad\qquad -C\,E\,F-C^*E^*F^*
\label{dett}
\eeqa
In this case, however, the inverse matrix (combined with the determinant) still
takes a simple form
\beq
\sqrt{\det \h}\,\h^{\abar b}= \left(
\begin{array}{ccc}
1+B& -C & -F^*\\
-C^* & 1+A & -E\\
-F & -E^* & 1+D 
\end{array} \right) \; .
\label{simp}
\eeq
For the D6-branes, we have $\calh_6=1+G$. Now the above
fields satisfy the low energy supergravity equations of motion if
$A,B,\dots, G$ are again harmonic functions on the transverse space.
Assuming that these harmonic functions all vanish asymptotically,
we would have 
\beq
A=\sum_{i=1}^n{a_i\over |\vec{x} -\vec{x}_i|}
\label{harmonic}
\eeq
and similarly for the other harmonic functions. From eq.~(\ref{simp}),
one easily sees that by setting $D=E=F=0$, one would recover the D-membrane
configuration of the previous section. However, the membranes would
now be delocalized in $y^5$ and $y^6$.

As in the six-dimensional solution, removing the
off-diagonal components of $\h_{a\bbar}$, by choosing $C=E=F=0$,
reproduces the standard solution describing a configuration
of orthogonal D-membranes and parallel D6-branes \cite{ortho}.
In fact as before, in the special case that all of the harmonic
functions have a single common center, our Hermitian metric solution
can be reduced to this known form by a coordinate transformation
on the world-volume coordinates, $z^a$. For generic harmonic functions,
however, our solution represents a generalization of the usual
harmonic superposition construction of extremal D-brane solutions \cite{ortho}.
One may extend the form (\ref{general6}) to the present 3$\times$3 Hermitian
matrix,
\beq
\h_{a\bbar}=\delta_{a\bbar}+\sum_{i=1}^n X_i\,\left[U_i\right]^1{}_a
\,\left[U_i^*\right]^1{}_\bbar
\label{general4}
\eeq
where the $X_i$ are harmonic functions on the transverse space,
and the $U_i$ are now SU(3) transformations acting on $z^{1,2,3}$.
With these choices, the Hermitian metric solution manifestly
describes D-membranes oriented at angles.

\subsection{Supersymmetry}

In this section, we show that the solution (\ref{solution}) saturates
the BPS bound, and hence that it preserves one-eighth of the
supersymmetries. 
In this calculation, we begin by computing the ADM mass
and the RR charge densities, using the appropriate
asymptotic flux integrals. These densities are then completely
determined by the leading-order asymptotic behavior of the harmonic
functions, which we define by
\beq
{1\over\sqrt{\det\h}}\,\h_{a\bbar}=\delta_{a\bbar}-{\hh_{a\bbar}\over r}
\label{asymp}
\eeq
where $\hh_{a \bbar}$ is a constant Hermitian matrix.
Also we set $G=g/r$, and we are using $r^2=(x^7)^2+(x^8)^2+(x^9)^2$.
Next, we examine
the Bogomol'nyi matrices of the individual constituents in order
to prove that the solution (\ref{solution}) 
preserves certain supersymmetries.  

The ADM mass per unit six-volume is defined by \cite{massy}:
\beq
m={1\over 2\ka^2} \oint \sum^{9}_{i=7} R^i\left[ \sum^{9}_{j=7} 
\partial_j \,\delta g_{ij} -
\sum^9_{j=1} \partial_i \,\delta g_{jj}\right] r^{2} d^2\Omega
\label{ADM mass}
\eeq
where $R^i$ is a radial unit vector in the transverse space, and 
$\delta g$ is the deviation of the asymptotic Einstein-frame metric $g^E$ from
flat space\footnote{The Einstein-frame metric is related to the string-frame
metric, which appears in the solutions, by $g^E_{\mu\nu} =e^{-{\phi\over 2}} 
g_{\mu\nu}$.}:
\beq
\delta g_{\mu\nu} = g^E_{\mu\nu}-\eta_{\mu\nu} \; .
\eeq
Using the asymptotic fields (\ref{asymp}) as described above,
the ADM mass density becomes
\beq
m ={2\pi\over\ka^2} (\Tr [h] + g) .
\label{mass}
\eeq
We note that this result is independent of the three (complex) constants
appearing in the off-diagonal components of $\hh_{a\bbar}$

The D-membranes produce asymptotic electric fields in the RR four-form
\ie $F^{(4)}_{tra\bbar}=-\prt_rA^{(3)}_{ta\bbar}$.
If we define (asymptotically orthonormal)
basis vectors, $n^a=\prt_{z^a}$ and their complex conjugates,
then we may define the associated
set of electric charge densities as\footnote{Note that this definition
is equivalent to that in \cite{angles}. There, the components
of $\qtil_{a\bbar}$ would be calculated by taking the
ten-dimensional Hodge dual of $F^{(4)}$, integrating the resulting fluxes
over the asymptotic two-sphere and four directions in the six-torus, but
then dividing out by the volume of the latter internal subspace.}
\beq
\qtil_{a\bbar}={1\over \sqrt{2} \ka} \oint *^{(4)}\left[i_{n^a}\,
i_{\nbar^b}\,F^{(4)}\right]
\label{electriccharge}
\eeq
where $*^{(4)}$ denotes Hodge duality on the noncompact four-dimensional
space, and $i_v$ denotes the usual interior product of vectors
with forms. The integral is performed over 
the asymptotic two-sphere in the transverse space. The simple final 
result is
\beq
\qtil_{a\bbar}=\pm{2\sqrt{2}\pi i\over\ka}\hh_{a\bbar}
\label{memdense}
\eeq
Note that since $i_{n^a}i_{n^b}F^{(4)}=0$, there are no fluxes or
charges associated with the (2,0) or (0,2) cycles of the internal
six-torus.

The D6-branes provide a magnetic RR charge defined by:
\beq
p ={1\over\sqrt{2}\ka} \oint F^{(2)} ={1\over\sqrt{2}\ka} \oint *F^{(8)}
\eeq
where $F^{(8)}=\dd A^{(7)}$ is the field strength of the 7-form potential of
the solution (\ref{solution}), and again the integral is performed over
the asymptotic two-sphere.  The resulting charge density is
\beq
p=\pm{2\sqrt{2}\pi\over \ka}\, g\ \ .
\label{magsix}
\eeq

To investigate the supersymmetries, we examine the Bogomol'nyi
matrix \cite{older,Pope}, which is
derived using both the supersymmetry algebra and the asymptotic form
of the background fields. The BPS bound is saturated and supersymmetries
are preserved when this matrix has eigenspinors with vanishing eigenvalue.
Following our conventions\footnote{Our ten-dimensional Dirac matrices
satisfy $\{\Ga_\mu,\Ga_\nu\}=-2g_{\mu\nu}$, and
$\Gamma_{10}=\Gamma_0\Gamma_1\Gamma_2\Gamma_3\Gamma_4\Gamma_5
\Gamma_6\Gamma_7\Gamma_8\Gamma_9$ so that $\left(\Gamma_{10}\right)^2=+
\one$. Also note that for the complex indices, we use
$\Ga_a=(\Ga_{y^{2a-1}}+i\Ga_{y^{2a}})/\sqrt{2}$ and 
$\Ga_\abar=(\Ga_{y^{2a-1}}-i\Ga_{y^{2a}})/\sqrt{2}$.}, 
one finds that the Bogomol'nyi
matrix $\calm$ is given by \cite{Pope}:
\beq
\calm = m\,\one + {i\over\sqrt{2}\ka}\,\qtil_{a\bbar}\ \Ga_0
\Ga_{[a}\Ga_{\bbar]}
-{i\over\sqrt{2}\ka}\,p\ \Ga_7\Ga_8\Ga_9\Ga_{10}
\label{totalm}
\eeq
where $\one$ denotes the identity matrix, and $[\ ]$ is used to
indicate antisymmetrization, as in $\Ga_{[a}\Ga_{\bbar]}=
\left(\Ga_{a}\Ga_{\bbar}-\Ga_{\bbar}\Ga_{a}\right)/2$.
Now unbroken supersymmetries will be signaled by the
presence of eigenspinors satisfying
$\calm\vareps=0$.

A more insightful approach is to consider pulling apart the individual
constituents of the general solution, \ie arrange the centers of the
independent harmonic functions far apart, and then examine the 
individual constraints that arise from each component. Hence we
begin by considering the solution (\ref{solution})
for which $g=0$ and only
$\hh_{1\bar{1}}$ is nonvanishing in eq.~(\ref{asymp}). This solution would
represent an isolated D-membrane lying in the ($y^1$,$y^2$) plane (and
also delocalized in the remaining internal directions). 
\beq
\calm = {2\pi\over\ka^2} \hh_{1\bar{1}}\left( \one \pm i\Gamma_0
\Ga_{y^1}\Ga_{y^2}\right) \; .
\eeq
The matrix enclosed in the brackets has precisely two distinct eigenvalues,
0 and 2. The eigenspinors with vanishing eigenvalue are those satisfying
\beq
\vareps_+=\mp i\Gamma_0\Ga_{y^1}\Ga_{y^2}\vareps_-
\label{dconst}
\eeq
where we have separated the two chiral components
$\vareps_\pm$ defined by:
\beq
\vareps_\pm = {1\over 2}\left(\one\pm\Gamma_{10}\right)\vareps\; .
\eeq
Thus we have recovered the well-known result that half of the
supersymmetries are preserved by a D-membrane, and the
eigenspinors satisfy precisely the constraint (\ref{dconst})
arising from the superstring worldsheet analysis of D-branes \cite{Polchin2}.
Similarly, the other diagonal cases where only $\hh_{2\bar{2}}\neq 0$ or
$\hh_{3\bar{3}}\neq 0$ lead to 
\beqa
\vareps_+=\mp i\Gamma_0\Ga_{y^3}\Ga_{y^4}\vareps_-
\label{againdconst}\\
\vareps_+=\mp i\Gamma_0\Ga_{y^5}\Ga_{y^6}\vareps_-
\label{moredconst}
\eeqa
respectively, as expected for a D-membrane lying in the ($y^3$,$y^4$)
or the ($y^5$,$y^6$) plane.
There are common solutions to these constraints 
(\ref{dconst}-\ref{moredconst}) leaving one-eighth of the 
supersymmetries unbroken \cite{ortho}.

The analysis of the case where only $g\neq0$, representing
an isolated D6-brane, is also straightforward.  In this case the 
Bogomol'nyi matrix is given by
\beqa
\calm &=& {2\pi\over\ka^2} g\left( \one \mp i\Gamma_{x^7}\Ga_{x^8}
\Ga_{x^9}\Ga_{10}\right)
\nonumber\\
&=& {2\pi\over\ka^2} g\left( \one \pm i
\Gamma_0\Gamma_{y^1}\Gamma_{y^2}\Gamma_{y^3}\Gamma_{y^4}\Gamma_{y^5}
\Gamma_{y^6}\right)
\; .
\label{sixcalm}
\eeqa
Here eigenspinors with vanishing eigenvalue satisfy the expected
D6-brane constraint:
\beq
\vareps_+=\mp i
\Gamma_0\Gamma_{y^1}\Gamma_{y^2}\Gamma_{y^3}\Gamma_{y^4}\Gamma_{y^5}
\Gamma_{y^6}
\vareps_- \; .
\label{susy2}
\eeq
It is also straightforward to show that the common eigenspinors satisfying
(\ref{dconst}-\ref{moredconst}) also satisfy the above constraint.

We must also consider some other configurations in which only off-diagonal
components of $\hh_{a\bbar}$ are nonvanishing. These are unusual since
from eq.~(\ref{mass}) one  finds that $m=0$ for these cases --- we will
comment more on these configurations in section \ref{discuss}.
Consider the solution with only $\hh_{1\bar{2}}=\hh_{2\bar{1}}\neq0$
in which case the Bogomol'nyi matrix reduces to
\beq
\calm = \mp i{2\pi\over\ka^2} \hh_{1\bar{2}}\,\Gamma_0\,
\left(\Ga_{y^1}\Ga_{y^4} + \Ga_{y^2}\Ga_{y^3}\right) \; .
\eeq
The appropriate eigenspinors must satisfy
\beq
\Ga_{y^1}\Ga_{y^4}\vareps_\pm=-\Ga_{y^2}\Ga_{y^3}\vareps_\pm
\label{yetmore}
\eeq
and it is easy to show that the common eigenspinors of eqs.~(\ref{dconst})
and (\ref{againdconst}) will satisfy this constraint. Similarly the
new constraints that are produced by considering the other
off-diagonal components of $\hh_{a\bbar}$ are already satisfied by
the common eigenspinors satisfying eqs.~(\ref{dconst}-\ref{moredconst}).
Therefore these contributions do not lead to the breaking of further
supersymmetries. Hence we arrive at the final conclusion that the
four-dimensional Hermitian solution (\ref{solution}) preserves
one-eighth of the supersymmetries.

\section{T-dual solutions}

The ten-dimensional T-duality
transformations between the type IIA and IIB superstring theories may
be found in ref.~\cite{Ortin} --- see also \cite{us}.
These transformations provide a powerful tool for generating
new solutions from a given background field configuration.
An important feature of T-duality is that it respects
supersymmetry. Therefore, all T-dual solutions of (\ref{solution}) will
also preserve one-eighth of the supersymmetries. In general, the T-dual
solutions will involve various non-threshold
D-brane bound states (as in \cite{us}),
but we will restrict ourselves to two cases which have
a simple interpretation as a marginally bound system of one or
two kinds of D-branes.

\subsection{4-4-4-0 configuration}

A particularly simple solution is obtained by T-dualizing all of
the six-torus coordinates in (\ref{solution}). The resulting
background field solution describes a system of D4- and D0-branes.
When the entire toroidally compactified space is
T-dualized, the sum of the internal metric with the 
Kalb-Ramond field is replaced by its inverse \cite{Giveon}: 
$g+B\rightarrow (g+B)^{-1}$.  In the present case without a Kalb-Ramond field
then, the internal metric is transformed to its inverse and so in the
final solution, the internal metric is again Hermitian.
The background fields of the T-dual 
solution may be written\footnote{Here, rather than writing the
RR field of the D4-branes as a magnetic three-form potential, we use
an electric five-form potential.}:
\beqa
ds^2 &=& \sqrt{\calh_0\calh_4}\left( -{dt^2\over \calh_0\calh_4} +2 
{\htil_{a\bbar}\, dz^ad\zbar^b \over \calh_4} +\sum^9_{i=7} (dx^i)^2 \right)
\nonumber\\
A^{(5)} &=& \pm\htil^{\abar a}\eps_{\abar\bbar\bar{c}}\eps_{abc}\; 
dt\wedge dz^b\wedge dz^c\wedge d\zbar^\bbar\wedge d\zbar^{\bar{c}}  \nonumber\\
A^{(1)} &=& \mp{1\over \calh_0}\; dt \label{solution2}\\
e^{2\phi} &=& {\calh_0^{3/ 2}\over \calh_4^{1/ 2}} \nonumber
\eeqa
where $\eps_{abc}$ is the three dimensional Levi-Civita symbol.
The components of the new hermitian matrix $\htil$ are related to the 
inverse of $H$ as
\beq
\htil_{a\bbar} = \sqrt{\det \h}\; \h^{\abar b} \; .
\label{hermitage}
\eeq
We have also introduced the notation: $\calh_0=1+G=\calh_6$, and 
$\calh_4=\det(\htil)=\sqrt{\det \h}=
\calh_2$, where the latter is explicitly given
in eq.~(\ref{dett}). If the off-diagonal components of $\htil$ are set
to zero, the above solution reduces to a marginally bound system
of orthogonal D4-branes and D0-branes, which was considered by 
\cite{balatwo} in investigating black hole entropy. We also remark
that if this approach of T-dualizing the entire compact space
were applied to the six-dimensional solutions of section 2, the
final result would again be in the same family of D-membrane solutions
presented there.

\subsection{3-3-3-3 configuration}

An interesting solution of the Type IIB theory is generated
by T-dualizing only half of the internal compact space.
If we restrict the hermitian matrix $\h$ in eq.~(\ref{solution})
to be real, one can construct a solution composed entirely of 
D3-branes. Hence after making this restriction, we dualize
on the internal coordinates $\{y^2,y^4,y^6\}$ to
obtain
\beqa
ds^2 &=& \sqrt{\calh\,\calh'} \left({-dt^2 +H_{ab}dy^{2a-1} dy^{2b-1}\over
\calh\,\calh'}+
{\htil_{ab}dy^{2a} dy^{2b}\over \calh} +\sum^9_{i=7}(dx^i)^2\right) 
\nonumber\\
F^{(5)} &=& \mp{\partial_i \calh'\over {\calh'}^2}\; 
dt\wedge dy^1\wedge dy^3\wedge dy^5\wedge dx^i
\pm\eps_{ijk} \partial_i \calh'\; 
dy^2\wedge dy^4 \wedge dy^6\wedge dx^j\wedge dx^k\nonumber\\
&& \pm\partial_i \left({H_{ab}\over \calh}\right) 
\eps_{bcd}\; 
dt\wedge dy^{2a-1}\wedge dy^{2{c}} \wedge dy^{2{d}}\wedge dx^i \nonumber \\
&&\pm\eps_{ijk}\partial_i\left( \calh \h^{ab} \right) \eps_{bcd}\;
     dy^{2a}\wedge dy^{2c-1}\wedge dy^{2d-1}\wedge dx^j\wedge dx^k \; . 
\nonumber \\
e^{2\phi} &=& 1  \label{solution3}
\eeqa
Here, the (real symmetric) matrices, $\h_{ab}$ and $\htil_{ab}$,
are defined by eqs.~(\ref{hermit}) and (\ref{hermitage}), respectively,
with the restrictions that $C=C^*$, $E=E^*$ and $F=F^*$. We also
denote $\calh=\det{\htil}=\sqrt{\det{\h}}=\calh_2$ and $\calh'=1+G=\calh_6$.
If we set the off-diagonal components to zero, \ie $C=0=E=F$,
then the solution describes and orthogonal system of D3-branes
oriented in the $(y^1,y^3,y^5)$, $(y^1,y^4,y^6)$, $(y^2,y^3,y^6)$
and $(y^2,y^4,y^5)$ surfaces.
These solutions were first considered in refs.~\cite{balatwo,igort}.

\section{Discussion} \label{discuss}

In this paper, we have presented low energy background field solutions
describing D-membrane configurations which are characterized by a
Hermitian metric on the effective worldvolume directions.
The components of this metric 
are composed of independent harmonic functions on the transverse space.
Thus these solutions generalize the usual harmonic superposition rule
for the construction of D-brane bound states \cite{ortho}.
We have 
explicitly shown that these new configurations saturate the BPS bound, 
preserving one-eighth (one-quarter) of the supersymmetries
in four (six) dimensions. The six-dimensional
solution of section 2 generalizes the solution describing D-membranes
at angles presented in \cite{angles}. In the four-dimensional solution,
the harmonic functions may be chosen so that the solution manifestly
describes D-membranes oriented at SU(3) angles with respect to one
another. The latter is in accord with the analysis of \cite{Berkooz}.
Their worldsheet techniques can be used to show that the supersymmetry
restrictions imposed by D-membranes at SU(3) angles are compatible,
leaving one-eighth of the supersymmetries unbroken. In fact, the
latter analysis is essentially the same as that of the Bogomol'nyi
matrix in section 3.1.

Generically, the Hermitian metric solutions generalize the standard
harmonic superposition rule \cite{ortho}. When the off-diagonal
components of the metric vanish, it is easily seen that our solutions
reduce to the orthogonal D-membrane solutions constructed by the
latter methods. Further, however,
at special points in the moduli space --- in particular,
when all of the harmonic functions have a single common center
--- the new solutions may be reduced to the known orthogonal D-membrane
solutions by a unitary transformation of the internal coordinates.
Since physical quantities like the ADM mass 
and Ramond-Ramond charge densities do not depend on the short-range 
structure of the solution, even a multi-center solution will appear
to have the same properties as some single center solution, and
hence some system of orthogonal D-membranes. Hence the new solutions
and the standard orthogonal configurations would be
indistinguishable for a far field observer. Of course, the
difference between the orthogonal and the general Hermitian
solutions will be significant when the 
short-range physics is important.

The solutions constructed here and by the usual harmonic superposition rule
are completely delocalized in the internal directions. Thus even though
these solutions may have an interpretation in terms of, \eg orthogonal
D-membranes, they show no structure in the directions parallel to the
worldvolume of one or more of the D-branes. Finding solutions which
explicitly display all of the structure of such intersecting D-branes
is a difficult and unresolved problem --- see \cite{comp,angledrama2,gaunt}.
The possibility to separate the centers of the harmonic functions does
provide some insight into the short distance structure of these 
configurations. For
a given single center solution describing a configuration of intersecting
D-membranes, one can see from the preceding discussion that there
is not a unique way to resolve the short distance fields. In particular,
while one can separate the centers in accord with the standard orthogonal
solution, the Hermitian metric solution can provide a distinct
resolution of the short range fields in terms of D-membranes oriented
at angles. Thus in smearing out or delocalizing the solutions, one has
in fact lost more information about the short distance physics than might
have been previously expected. 

An interesting question is whether or not this non-uniqueness has any
implications for the black hole entropy calculations performed in
a D-brane framework. In fact, if one focuses on the single center
solutions the inclusion of D6-branes along with the D-membranes in 
the four-dimensional solutions of section 3 generically  
produces a nonsingular black hole solution. Must one then take
into account all possible orientations of D-membranes which might
describe a black hole with a given set of four-dimensional charges
in calculating the ground state degeneracy?
We expect that the answer is no, to leading order.
While the above discussion indicates that here and perhaps in other cases, there
may be more configurations than considered at first sight, this will
be a subleading contribution, which does not affect the dominant exponential
degeneracy. It would be rather like
considering contributions of the D-string winding states in
the standard five-dimensional calculations \cite{blackholes}. In principle,
all of these configurations make distinct contributions in any regime
of interest, however, it is sufficient to focus on a single winding
configuration to determine the leading order
degeneracy \cite{blackholes,fat}. Still the
non-uniqueness discovered here would be important if one wishes to make
a detailed comparison of subleading contributions to the black hole
entropy.

In examining the Bogomol'nyi matrices in section 3.1, some interesting
configurations were considered for which the mass vanished.
For concreteness, let us consider the six-dimensional solution
(\ref{oldsolution}) with $A=0=B$ and $C=C^*=c/|\vec{x}|^3$. It is
straightforward to show that this solution has a vanishing mass,
but nonvanishing charges. The solution is singular at $C=1$ where
$\calh=0$ --- the presence of the singularity can be determined
by examining the curvature, or more easily by considering simpler
invariants such as $(\nabla\phi)^2$. At this point, the Hermitian
metric is degenerate as the $y^1$ and $y^3$ directions, as well as
$y^2$ and $y^4$, become collinear. Note however that in the string
frame metric, the proper volume of the internal four-torus remains
finite as $C\rightarrow1$ because of the factors of $\calh^{-1/2}$
in $G_{a\bbar}$. Since this example is a single center solution,
one can consider which orthogonal solution results upon making
a coordinate transformation (\ref{diag}). In this case, the
transformed diagonal metric has the form
\beq
H=
\left(
\begin{array}{cc}
1+{c\over|\vec{x}|^3} & 0 \\
0 & 1-{c\over|\vec{x}|^3}
\end{array}
\right)\ \ .
\eeq
Hence the corresponding configuration of two orthogonal D-membranes
has a vanishing mass because the mass density of one of the constituents
is actually negative! This unsavory feature may indicate
that these solutions should not be regarded as physical.
Finally, we note that these curious configurations
are reminiscent of massless black hole solutions found
in the low energy heterotic string theory \cite{nomass}.

It would be of interest to more fully examine the family
of new solutions which might be generated from those presented
here by means of T-duality. Raising these Type IIA supergravity
solutions to eleven-dimensional solutions would also be an interesting
avenue of exploration. Finally we note that there would be no 
new (stationary) nonextremal solutions.
In a nonextremal configuration,
the branes would collapse forming a black hole at a particular
position. Thus the final configuration would be described
by some single center solution. However, one should expect that
as for the extremal solutions these nonextremal single center solutions
will be equivalent up to a coordinate transformation to the standard
solutions \cite{none}, which are interpreted in terms of orthogonal branes.

\section*{Acknowledgments}
We gratefully acknowledge useful conversations with Jason Breckenridge,
Don Marolf and Arkady Tseyt\-lin.
This research was supported by NSERC of Canada and Fonds FCAR du
Qu\'ebec.

\end{document}